\documentclass{elsart}
\usepackage{times}
\usepackage{graphicx}
\begin{document}

\begin{frontmatter}

\title{Design Study of CANGAROO-III, Stereoscopic Imaging
Atmospheric Cherenkov Telescopes for sub-TeV Gamma-ray 
Detection\thanksref{submit}}
\thanks[submit]{to be appeared in Astroparticle Physics.}

\author[icrr]{R. Enomoto\thanksref{email1}}, 
\thanks[email1]{enomoto@icrr.u-tokyo.ac.jp}
\author[tit]{S. Hara\thanksref{email2}},
\thanks[email2]{shara@hp.phys.titech.ac.jp}
\author[tit]{A. Asahara},
\author[anu]{G.V. Bicknell},
\author[isas]{P.G. Edwards},
\author[yamagata]{S. Gunji},
\author[yamanashi]{T. Hara},
\author[tokai]{J. Jimbo},
\author[konan]{F. Kajino},
\author[icrr]{H. Katagiri},
\author[kyoto]{J. Kataoka},
\author[icrr]{A. Kawachi},
\author[icrr]{T. Kifune},
\author[kyoto]{H. Kubo},
\author[kyoto]{J. Kushida},
\author[ste]{Y. Matsubara},
\author[nao]{Y. Mizumoto},
\author[icrr]{M. Mori},
\author[tit]{M. Moriya},
\author[ipu]{H. Muraishi},
\author[ste]{Y. Muraki},
\author[yamanashi]{T. Naito},
\author[tokai]{T. Nakase},
\author[tokai]{K. Nishijima},
\author[icrr]{K. Okumura},
\author[adelaide]{J.R. Patterson},
\author[tit]{K. Sakurazawa},
\author[adelaide]{D.L. Swaby},
\author[tit]{K. Takano},
\author[kyoto]{T. Tanimori},
\author[kanagawa]{T. Tamura},
\author[icrr]{K. Tsuchiya},
\author[tokai]{K. Uruma},
\author[ibaraki]{S. Yanagita},
\author[ibaraki]{T. Yoshida},
\author[ocu]{T. Yoshikoshi},
\author[ste]{A. Yuki}

\address[icrr]{ICRR, Univ. of Tokyo, 5-1-5 Kashiwa-no-ha, 
Chiba 277-8582, Japan}
\address[tit]{Tokyo Inst. of Tech., Meguro-ku, Tokyo 152-8551, Japan}
\address[anu]{Australian National University, ACT 2611, Australia}
\address[adelaide]{Dept. of Phys. and Math. Phys., Univ. Adelaide, SA 5005, 
Australia}
\address[isas]{ISAS, Sagamihara, Kanagawa 229-8510, Japan}
\address[yamagata]{Dept. Phys., Yamagata Univ., Yamagata, Yamagata 990-8560, 
Japan}
\address[ibaraki]{Ibaraki University, Mito, Ibaraki 310-8512, Japan}
\address[yamanashi]{Yamanashi Gakuin University, Kofu, Yamanashi 400-8575, 
Japan}
\address[konan]{Dept. of Phys., Konan University, Kobe, Hyogo 658-8501, Japan}
\address[tokai]{Tokai University, Hiratsuka, Kanagawa 259-1292, Japan}
\address[kyoto]{Dept. of Phys., Kyoto University, Sakyo-ku, Kyoto 606-8502, 
Japan}
\address[ste]{STE Lab., Nagoya University, Nagoya, Aichi 464-8602, Japan}
\address[nao]{National Astronomical Observatory, Mitaka, Tokyo 181-8588, Japan}
\address[ipu]{IPU, Ami, Ibaraki 300-0394, Japan}
\address[kanagawa]{Kanagawa University, Yokohama, Kanagawa 221-8686, Japan}
\address[ocu]{Dept. of Phys., OCU, Osaka, Osaka 558-8585, Japan}

\begin{abstract}
CANGAROO-III is an Imaging Atmospheric Cherenkov Telescope
(IACT) array of four 10\,m telescopes for
very high energy (sub-TeV) gamma-ray astronomy.
A design study of the CANGAROO-III telescope system was carried out using 
the Monte Carlo technique in order to optimize
the pixel size and the telescope spacing.
Studies were also made of observations at low elevation angles. 
\end{abstract}


\end{frontmatter}

\section{Introduction}

CANGAROO-III is an Imaging Atmospheric Cherenkov Telescope
(IACT) array observing very high-energy (sub-TeV) gamma-rays
from the universe. The four 10\,m diameter 
telescopes can operate independently or as an array. 

The CANGAROO experiment is located in Woomera, South Australia.
It started from a 3.8-m telescope\cite{ref:nim} (CANGAROO-I) in 1992.
The second stage, CANGAROO-II, commenced in 1999 with the construction
of a 10\,m telescope mount with a mirror initially of 7\,m diameter.
The mirror was extended to 10\,m in 2000 \cite{ref:c2}.
CANGAROO-III will be an array of four telescopes, with the existing 
CANGAROO-II telescope being the first of these.

We studied the basic design parameters for the next three telescopes
using the Monte Carlo technique.
Similar studies have been carried out previously, for example by 
the VERITAS \cite{ref:veritas} and HESS \cite{ref:hofmann} groups.
We started from optimized parameters obtained by a previous
study and our experience with CANGAROO-I,\cite{ref:nim} and verified
those results. 
The pixel size and the telescope spacing were optimized.
Also, some studies concerning the geomagnetic
field effects and the possibility of large zenith angle 
observations were included.

\section{Monte Carlo}

Electromagnetic and hadronic shower simulations in
the air using a Monte-Carlo code were conducted based on 
GEANT3.21 \cite{ref:geant321}.
In this method, the atmosphere is divided into 80 layers
of equal thickness ($\sim 12.9g/cm^2$). 
Each layer corresponds to less than a half radiation length.
The dependence of results on the number of layers was checked by 
halving the number of layers and was confirmed to be less than a 10\% effect.
The lower energy threshold for particle transport was set at
20\,MeV, which is less than the Cherenkov threshold of electrons
at normal temperature and pressure (NTP).
Most Cherenkov light is emitted higher in the atmosphere, i.e., 
at less pressure and a higher Cherenkov threshold.
The geomagnetic field at the Woomera site (South Australia) was 
included in the simulations (0.253 gauss in horizontal and 0.520 gauss
in vertical directions, and 6.8$^\circ$ degrees off from south).

In order to save CPU time,
Cherenkov light was tracked in the simulations only when it was
initially directed to the mirror area.
The average mirror reflectivity and
the measured photo-multiplier (PMT) quantum efficiency were 
multiplied by the Frank-Tamm equation
to derive the total amount of light and its wavelength dependence. 
A Rayleigh-scattering length
of $2970(\lambda/400nm)^4$ $[g/cm^2]$ was used in transport to the
ground \cite{ref:rayleigh}. 
No Mie scattering was assumed in this study.
The contribution of Mie scattering is thought to be at the 10--20\% level,
and we therefore consider this study to have uncertainties at at least
this level.
When Rayleigh scattering occurred, we treated it as absorption.
The diameter of each spherical mirror segment is 80\,cm. 
We assumed a perfect spherical shape (i.e., no deformation and
no blur spot) with the CANGAROO-II geometry \cite{ref:mirror}.
The main telescope parameters are listed in Table \ref{tab:cII}.
\begin{table}
  \caption{\it Main parameters of CANGAROO-II telescope.}
  \begin{tabular}{ll}
    \hline
    \hline
    parameters & values \\
    \hline
    coordinate & 136$^\circ$E, 31$^\circ$S \\
    height above sea level & 220m \\
    total diameter & 10m \\
    focal length & 8m \\
    number of mirror-segmentation & 114 \\
    mirror radius & 80cm \\
    mirror shape & spherical \\
    mirror alignment & parabolic \\
    mirror curvature & 16.4m \\
    mirror material & plastic \\
    \hline
    \hline
  \end{tabular}
  \label{tab:cII}
\end{table}
The average measured reflectivity of 80\% at 400\,nm was adopted.
The curvature of the mirror segments is 16.4\,m.
In total, 
114 mirrors were aligned in a parabolic shape with a focal length
of 8\,m.\cite{ref:mirror}

For the simulations, we tried the use of four different camera designs:
\begin{description}
\item[setup~1:] Five hundred and seventy-six 1/2$''$ PMTs of the same 
type as the CANGAROO-II camera \cite{ref:c2}.
With a pixel spacing of 0.112$^\circ$, the
total field of view (FOV) was $2.7^\circ\times 2.7^\circ$ square.

\item[setup~2:] Five hundred and seventy-six 3/4$''$ PMTs. 
A pixel spacing of 0.168$^\circ$ yields a FOV of 4$^\circ$ square. 

\item[setup~3:] Two hundred and fifty-six 3/4$''$ PMTs, with
a FOV of 2.7$^\circ$ square.

\item[setup~4:] Five hundred and seventy-six 3/4$''$ PMTs 
with a smaller PMT separation than in setup~2:
a pixel spacing of 0.147$^\circ$ and a FOV of 3.5$^\circ$.
\end{description}
The shape of the photocathodes was assumed to be circular.
Each PMT had a light guide above the surface of the photo-cathode. 
The gain in light yield was assumed to 
be 1.6 compared to the case without it.\cite{ref:c2}

In order to simulate cosmic-ray background events,
we generated only protons with a differential energy spectrum of $E^{-2.7}$.
The minimum and maximum energies of the generated range were
100 and 5000\,GeV, respectively, in the case of zenith injection.
For the large zenith angle study, we increased the maximum
energy to 20 TeV.
The maximum core distance of simulated showers was 300\,m in radius
for the zenith injection. We also increased it for the large angle
injection. For example, we used 1000\,m for 55$^\circ$ injection.
The maximum offset angle for cosmic-ray showers was 5~degrees.
We generated typically 100,000 events for each setup.

For gamma-ray cascades, we chose $E^{-2.5}$ spectrum (Crab-like spectrum
\cite{ref:largeangle}).
The minimum and maximum of the generated energies were 50 and 5000\,GeV,
respectively, for cascades from the zenith, with the maximum increased
for the low elevation simulations.

Finally, electronics noise was added and the timing responses
were smeared using Gaussian of 4 nsec ($\sigma$). 
We also added Night Sky Background (NSB) photons,
conservatively selecting to double Jelley's value 
of $2.55 \times 10^{-4} erg/cm^2/sec/sr$ (430-550 nm) \cite{ref:Jelley}.

\section{Analysis}

\begin{figure}
\includegraphics[width=11cm]{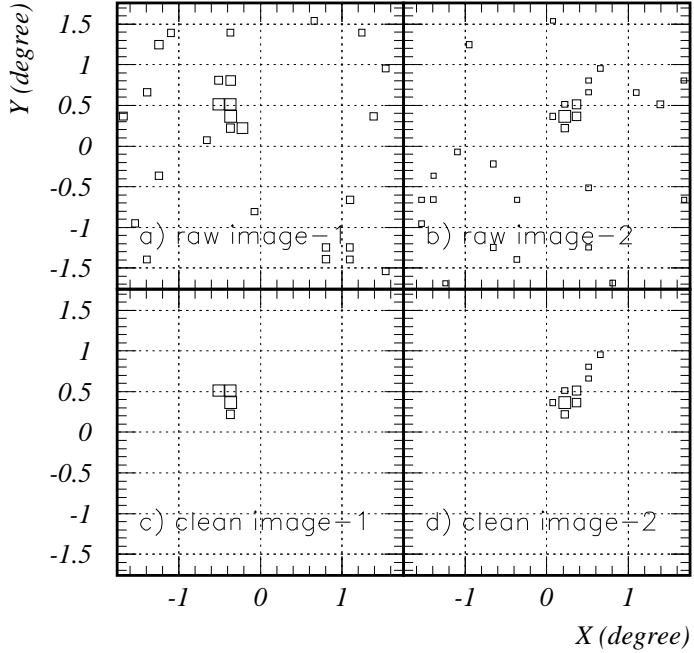}
\caption{\it Simulated camera images for two telescopes separated by 100\,m. 
The box sizes are proportional to the number of
photo-electrons. a) raw image of telescope~1, b)
raw image of telescope~2, c) cleaned image of telescope~1, and
d) cleaned image of telescope~2.
}
\label{cluster}
\end{figure}

First, we applied a threshold for the PMT pulse-heights. The threshold
was set at 5~photoelectrons. This greatly reduced the effect of
NSB photons.
Second, we applied a clustering cut. 
Only PMTs exceeding the threshold (``hits'') and having more than two adjacent
hits were selected (t3a clustering\cite{ref:veritas}).
A hit map of an event is shown in Figure \ref{cluster}.

The left figures are for telescope-1 and the right are for telescope-2.
The spacing of the telescopes was 100\,m. 
The box sizes are proportional to the number of photoelectrons.
The upper figures were obtained
without this clustering and the lower ones used this clustering cut.
Isolated pixels triggered by NSB photons were all removed by this operation.

\begin{figure}
\includegraphics[width=11.0cm]{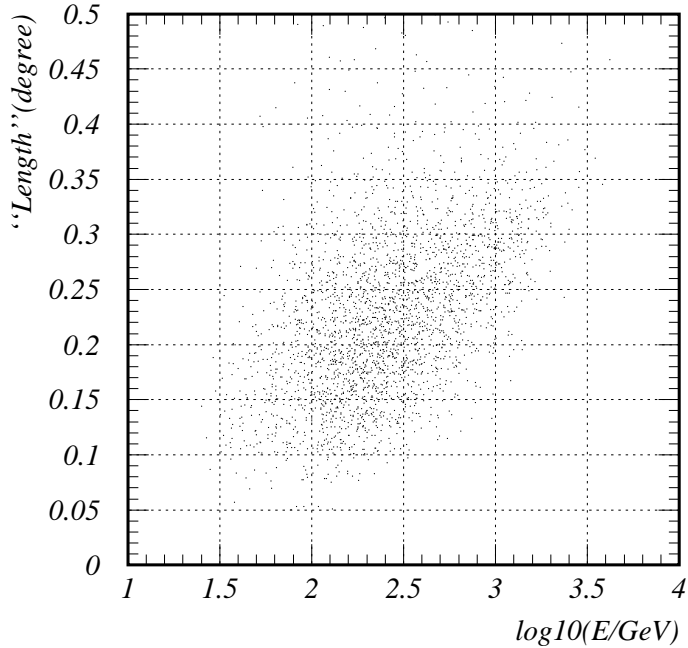}
\caption{\it Energy dependence of the ``length" parameter, i.e., 
the ``length"
versus the detected number of photo-electrons.}
\label{edeplen}
\end{figure}

The selection of gamma-ray events was carried out using
the standard imaging analysis technique. 
The imaging analysis was based on 
parameterization of the Cherenkov light
image by its ``width," ``length," ``conc" (shape), 
``distance" (location),
and the image orientation angle ($\alpha$) \cite{ref:shape}.
The energy dependence of some parameters is shown in Figure~\ref{edeplen}.
For example, ``length" has a linear correlation with the logarithm of energy.
We, therefore, used a linearly corrected ``length" instead of the original
one. We applied similar corrections to the other parameters.

\begin{figure}
\includegraphics[width=11.0cm]{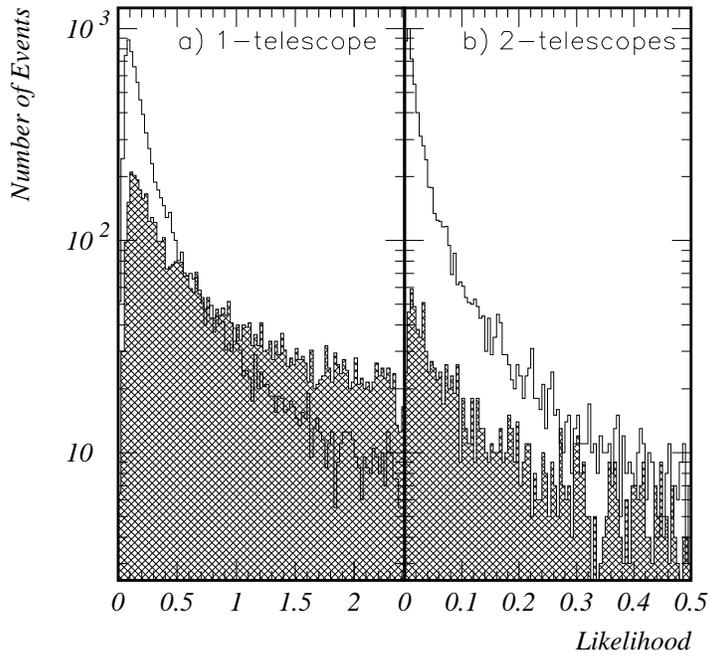}
\caption{\it Log-Likelihood distribution, i.e., $\chi^2$. The blank histograms
are for gamma-rays and the hatched areas are protons; a) for 1 telescope,
and b) for 2 telescopes (in stereo mode).}
\label{like}
\end{figure}

Using these parameters, except for $\alpha$, we carried out
a likelihood analysis in order to separate gamma-ray--like showers
from hadron-like showers \cite{ref:like}.
In order to make a probability density function (PDF), we generated
two kinds of Monte Carlo event samples. The gamma-ray sample was
made assuming the primary flux to be proportional to $E^{-2.5}$
(the spectrum of the Crab pulsar \cite{ref:largeangle}).
The hadron sample assumed a spectrum of $E^{-2.7}$.
Using these two samples, we made PDFs from histograms of the shower
parameters. We calculated the probability
on an event-by-event basis of
the event being due to a gamma-ray using
$$Prob=\Pi \{Prob(i,\gamma)/(Prob(i,\gamma)+Prob(i,p))\},$$
where the suffix i refers to the shower parameters.
A cut was subjected to the parameter $\chi^2$s, which is
$-2\ln(Prob)$, as shown in Figure \ref{like}.
The smaller $\chi^2$ events are likely to be gamma-ray events.
The blank histograms are for gamma-ray events and the hatched one for
hadrons. The left hand figure is for a single-telescope analysis, and the right
one is for a stereo (two telescope) analysis.
In analyzing the stereo case, the probability was calculated
by $Prob(telescope-1)\times Prob(telescope-2)$.
The signal-to-noise ratios (S/N) were greatly improved by stereoscopic
observations while maintaining shower detection efficiency
when both telescopes have analyzable images.

\section{Pixel Size}
\begin{figure}
\includegraphics[width=11.0cm]{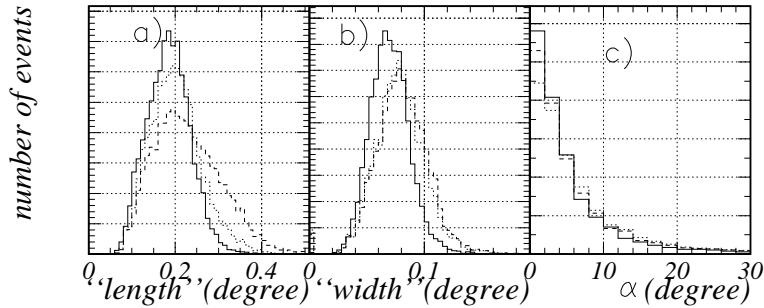}
\caption{\it Comparison of parameters (a) length, (b) width, and
(c) $\alpha$ for three camera designs. 
The curves are for gamma-rays.
The solid lines are for setup~1, 
the dashed lines are setup~2,
and the dotted lines are setup~3.}
\label{comp}
\end{figure}

The present pixel spacing of the CANGAROO-II 
camera is 0.112$^\circ$ \cite{ref:c2}.
This is based on the use of 1/2$''$ PMTs and is the
setup~1 described  previously.
This setup has a FOV of 2.7$^\circ$ square. With this FOV,
we are forced to conduct so-called ``Long ON/OFF" mode
for observations \cite{ref:psr1706}. 
For example, a 3\,hr ``ON" source observation and a 3\,hr ``OFF" 
source observation may be made during one night.
This operation reduces the observation period
significantly --- by a factor of two.
In order to carry out ON and OFF source observations simultaneously, 
we need a FOV of approximately 4$^\circ$. 
As the major part of shower images for gamma-rays are  typically
contained within a 1$^\circ$ circle,
we investigated the use of setup~2, with a larger PMT size 
(3/4$''$ circular photocathode).
Because of the weight limitation due to the telescope structure,
we can not increase the camera weight very much \cite{ref:mirror}.
We compared the shower image parameters for both cases, as shown in
Figure~\ref{comp}.
The deterioration of $\alpha$ resolution is very small, as shown in 
Figure~\ref{comp}c). The width deteriorates slightly due to the pixel-size
effect, but not that greatly (Figure \ref{comp}b).
The length, however, changed significantly (Figure~\ref{comp}a).
In order to check the FOV effect, we tried setup~3, i.e., 
the same PMT size as setup~2 but the same FOV as setup~1.
From Figure~\ref{comp}a it is apparent that 
the deterioration in length is due to the change in the FOV.
We conclude that the smaller FOV deforms the length distribution
and that setup~2 is preferred.

\begin{figure}
\includegraphics[width=11.0cm]{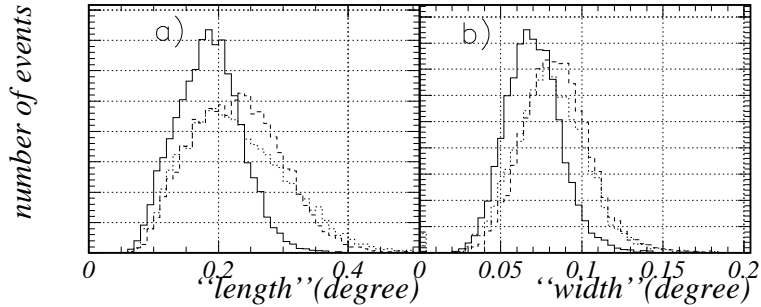}
\caption{\it Comparisons between setup~4 and setups~1 and 2 for  (a) length,
and (b) width: The solid line is setup~1, the dotted setup~2,
 and the dashed setup~4. The curves are for gamma-rays.}
\label{comp2}
\end{figure}

A 3/4$''$ PMT has a diameter of $18.6\pm 0.7$\,mm\cite{ref:hpkk}.
In setup~2, the spacing of the PMTs was 24\,mm. A clearance
of 5\,mm was kept. We tried setup~4 with a spacing
of 21\,mm (clearance of 2\,mm). 
Comparisons in Figure~\ref{comp2}
show that the difference is sufficiently small.
We again concluded that the setup~2 is a reasonable choice.
In the real design of the CANGAROO-III camera, we have selected
a ``hexagonal" arrangement of PMTs with alternate rows
offset by half a pixel. 

\section{Telescope Spacing}

\subsection{Stereo Mode}

\begin{figure}
\includegraphics[width=11.0cm]{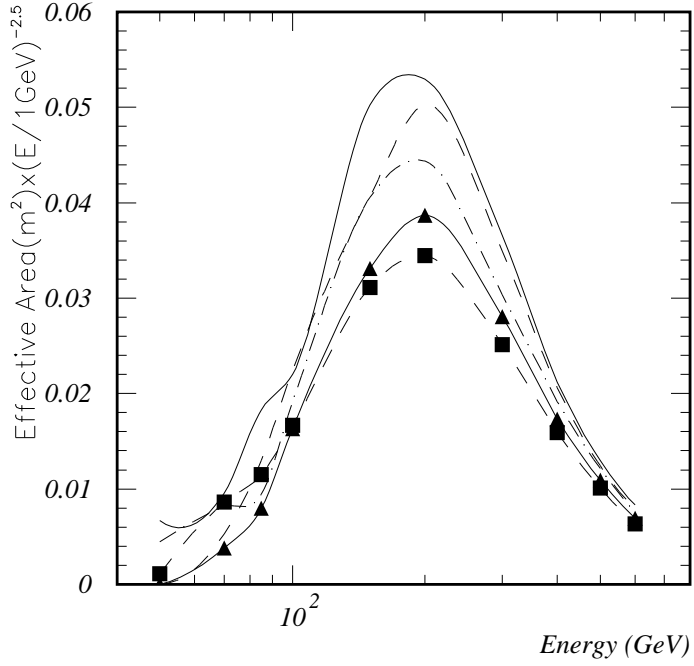}
\caption{\it ``Effective area" versus the incident $\gamma$-ray energy.
The vertical axis is the effective area ($m^2$) multiplied by
the $(E/GeV)^{-2.5}$-energy spectrum. 
The telescope spacing was varied from 60 to 140\,m;
the solid line is for 60\,m, the dashed 80\,m, the dot-dashed 100\,m,
the solid with triangles 120\,m, and the dashed with squares 140\, m.
}
\label{seff}
\end{figure}

With a likelihood analysis and the above-mentioned camera design (``setup-2"),
we proceeded to design a stereoscopic telescope system.
Previous studies have investigated the effect of
the telescope spacing \cite{ref:veritas}\cite{ref:hofmann}.
These indicated that a spacing within a range of 80 to 120\,m
is best. We, therefore, tested the range between 60 and 140\,m.
Here, we assumed two telescopes.
We started from calculations of ``effective area"  versus the incident
gamma-ray energy.
Gamma-rays of various energies from the zenith were generated.
The results are shown in Figure \ref{seff}.
Here, the effective area is the product of the ``real" effective 
area ($m^2$) and
the Crab-like spectrum $(E/GeV)^{-2.5}$ in order to show the effective
threshold. 
Those curves were obtained after smoothing.
They peak at around 200\,GeV and are a decreasing function of the telescope
spacing. The light pool on the surface has a radius of approximately
100\,m. The coincidence rate of the two telescopes, therefore, decreases
as the spacing increases.

\begin{figure}
\includegraphics[width=11.0cm]{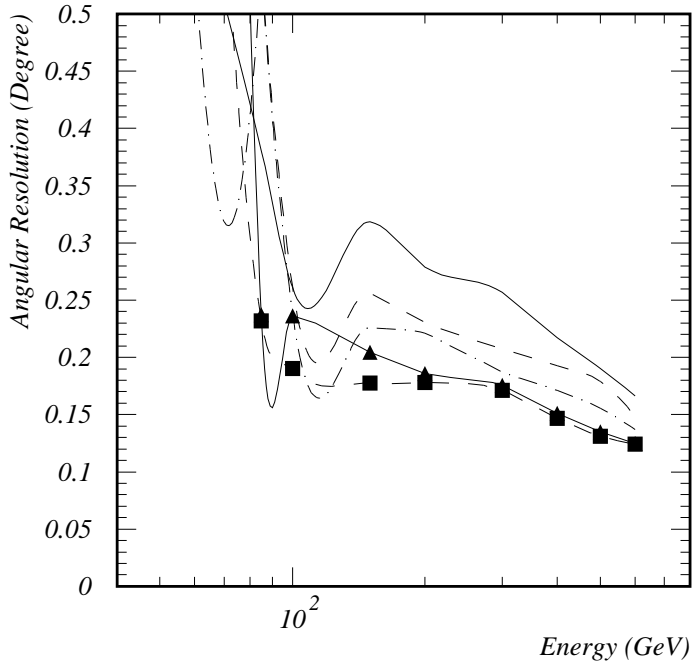}
\caption{\it Angular resolution versus the incident $\gamma$-ray energy.
The resolution is obtained on an event-by-event basis.
The solid line is for 60\,m, the dashed 80\,m, the dot-dashed 100\,m,
the solid with triangles 120\,m, and the dashed with squares 140\,m.
}
\label{sang}
\end{figure}

In the stereoscopic mode, the images from the two telescopes should
point to the same direction, i.e., to the source direction.
The angular resolution of this method should improve
with a larger telescope spacing due to the larger opening angles of images.
This is shown in Figure~\ref{sang}. 
In the figure, the curves ``wiggle'' below 100\,GeV
due to the lack of Monte-Carlo statistics, but the general
trend of worsening angular resolution with decreasing energy is clear.
This has an opposite energy dependence compared with that 
of the effective area.
Typically, at around the threshold ($\sim$200 GeV), the angular resolution
is 0.2$^\circ$ per shower.

\begin{figure}
\includegraphics[width=11.0cm]{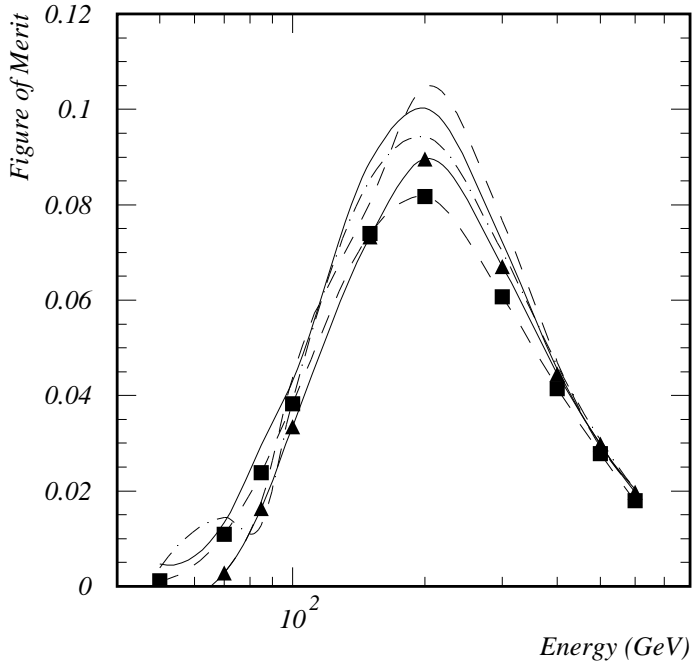}
\caption{\it Figure of merit (described in the text) versus the incident
$\gamma$-rays energy.
The solid line is 60\,m, the dashed 80\,m, the dot-dashed 100\,m,
the solid line with triangles 120\,m, and the dashed with squares 140\,m.
}
\label{fmerit}
\end{figure}

For a point-source observation, 
we can define the following figure of merit (FOM)
using the above two parameters:
$$FOM=``effective~area"/\sqrt{angular~resolution}
[m^2GeV^{-2.5}(degree)^{0.5}].$$
This value is proportional to the statistical significance of 
the observations.
The energy dependence of the FOM for various telescope spacings 
are plotted in Figure~\ref{fmerit}.
The FOM is maximized at the 80\,m spacing. The dependency on the spacing,
however, is small within this range. Also, the energy threshold for
the stereoscopic mode was obtained to be 200\,GeV.
In conclusion, telescope spacings of between 60 and 140 m are all
acceptable.
We selected a 95--100\,m spacing for the CANGAROO-III experiment.

\subsection{Monocular Mode}

\begin{figure}
\includegraphics[width=11.0cm]{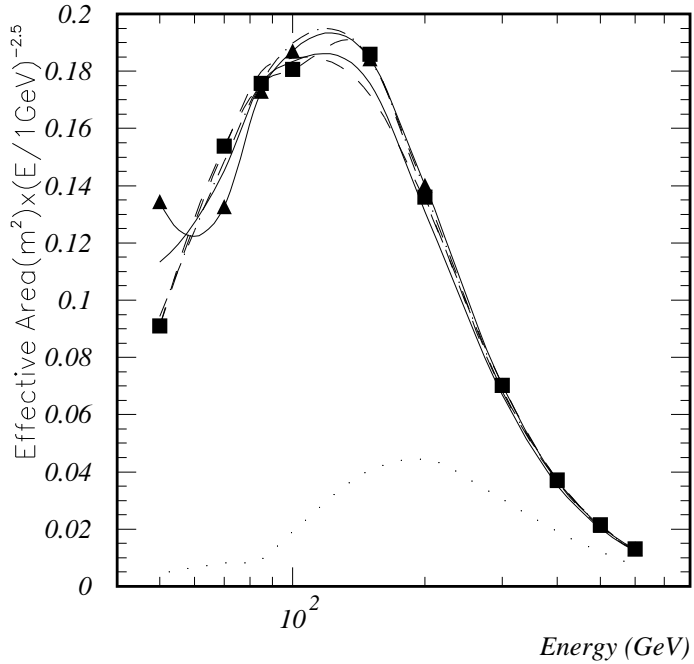}
\caption{\it Total detection areas versus incident gamma-ray energy
with ``monocular mode",
i.e., by an independent two-telescope analysis.
The solid line is 60\,m, the dashed 80\,m, the dot-dashed 100\,m,
the solid line with triangles 120\,m, and the dashed with squares 140\,m.
The dotted line is the effective area obtained for the two-telescopes
configuration (stereo mode) with 100m separation.
}
\label{sor}
\end{figure}

We can operate multi-telescopes independently. 
We assumed that data from each telescope were recorded whenever it triggers,
and that ``stereo" triggering is done later in software analysis.
We call this the ``Monocular Mode". 
In this mode, when all telescopes have the same pointing direction,
we expected a much improved effective area, especially at low energies.
The calculated effective area for the monocular mode
is shown in Figure \ref{sor}.
It was obtained under the trigger condition, in which one of two telescopes
satisfies the analysis conditions.
The gain in the effective area at lower energies is very much
improved. The energy threshold of this mode was obtained to be
100\,GeV. We hope that this mode will work as a ``discovery mode".
At  higher energies, such as 1\,TeV, the effective areas for both
the stereo and monocular modes coincide.

\subsection{Large Angle Observation}

\begin{figure}
\includegraphics[width=11.0cm]{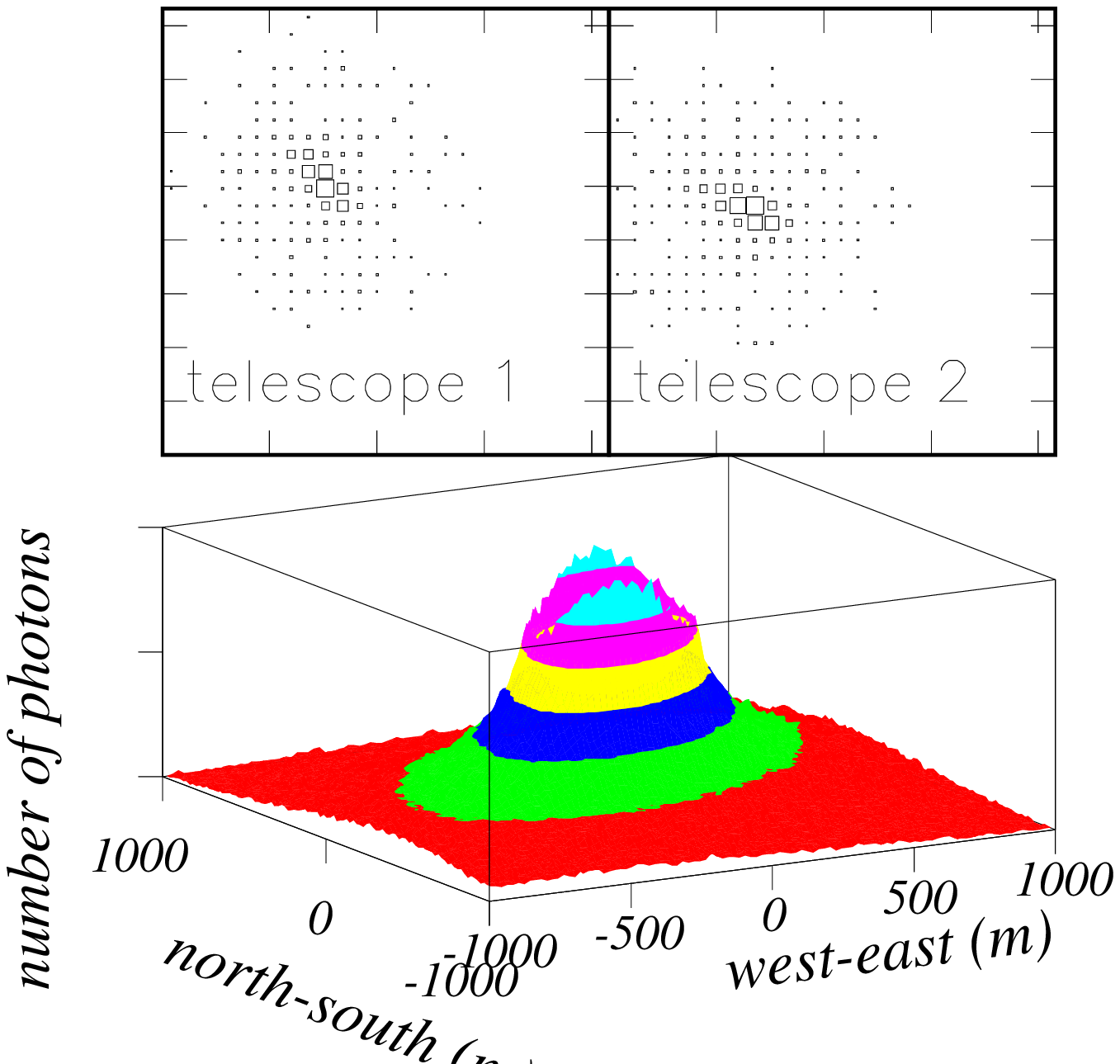}
\caption{\it Light pool of a large zenith angle event (lower plot). 
The scale is in meters.
The upper plots are the focal-plane images for two telescopes.}
\label{crab}
\end{figure}

It is well known that observations at large zenith angles
(i.e., low elevation angles) are effective for higher energy
measurements \cite{ref:largeangle}.
This is due to an increase in an effective area because of the inclined
injection of gamma-rays to the atmosphere.
For example, we show the light pool on the Earth's surface of a
gamma-ray shower which is injected 55$^\circ$ inclined from the zenith
(Figure~\ref{crab} lower plot). The azimuthal direction
of this injection is north-west.
The camera of the setup-2 was used.
Here, we also used an 80-layer atmosphere and the GEANT package.
Gamma-rays from the zenith make light pools
with the density peaking at a radius of 120\,m.
The light profile
is enlarged by a factor of $\sim$4, 
most notably in the longitudinal direction but also
in the transverse direction.
The upper plots are the focal-plane images of this shower.
The drawback with this technique is that shower images shrink.
The gamma/hadron discrimination, therefore, using shower parameters,
deteriorates, reducing the S/N ratio and the 
$\alpha$ resolution is also worsened. 
Although the detection possibility for low elevation observing
was proved by CANGAROO-I
in observations of the Crab \cite{ref:largeangle},
we can justify the use of this technique even for stereoscopic modes.
The differences in the shape parameter distributions are shown in 
Figure~\ref{image}:
\begin{figure}
\includegraphics[width=11.0cm]{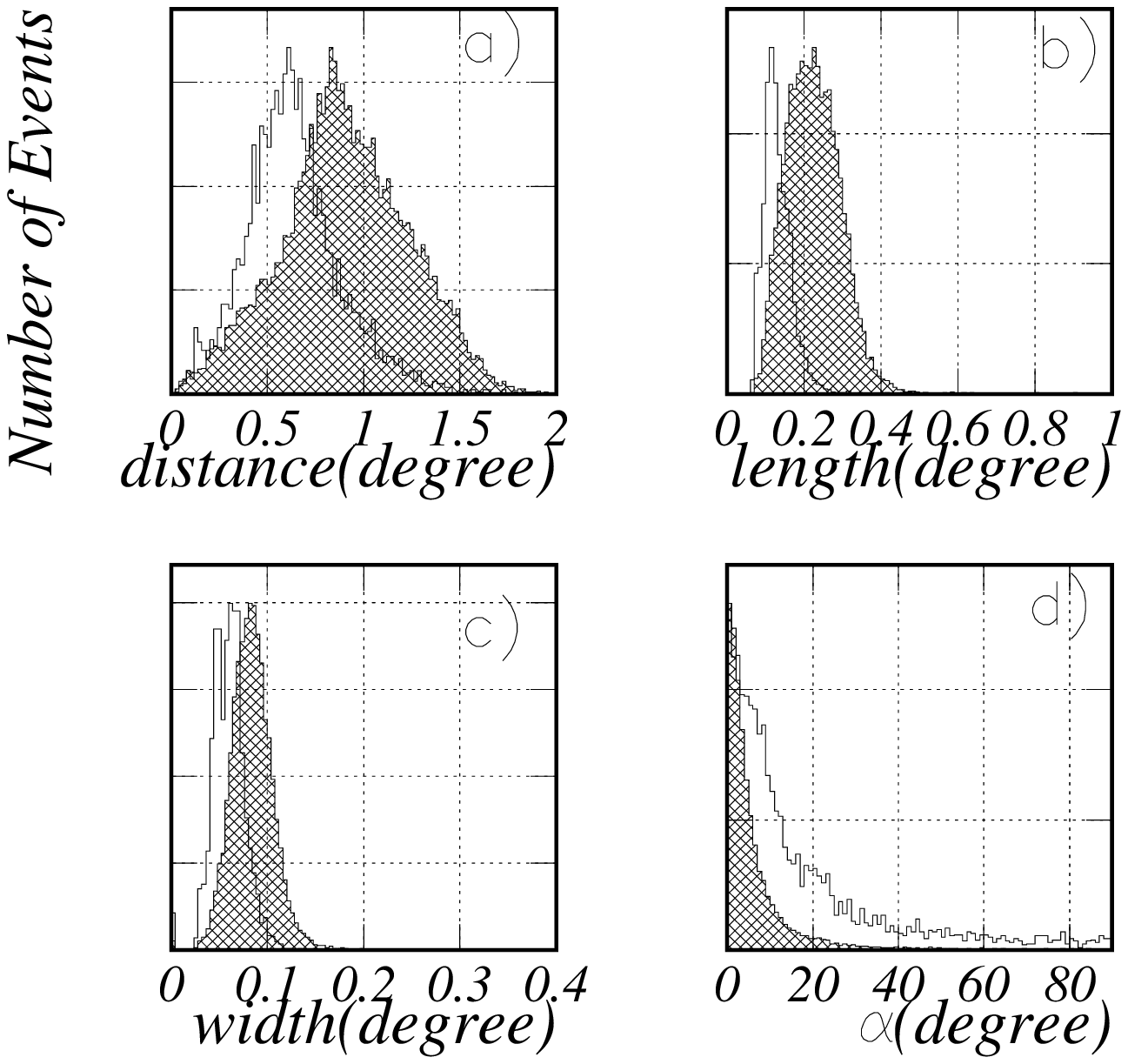}
\caption{\it 
Comparisons of shower parameters: a) ``distance", b) ``length", c) ``width",
and d) $\alpha$. The blank histograms were obtained by 55$^\circ$-observations
and the hatched ones by zenith observations.
The shower parameters were shrunk due to the large distances between the
telescope and the shower-max positions.
}
\label{image}
\end{figure}
for distance, length, width, and $\alpha$.
The $\alpha$ distribution still has a peak, as shown in Figure~\ref{image}d.
The other parameters also have differences between gamma-rays
and hadrons.
Although the gamma-ray images shrink,
the hadron images shrink in the same way and it is 
still possible to discriminate between them.

\begin{figure}
\includegraphics[width=11.0cm]{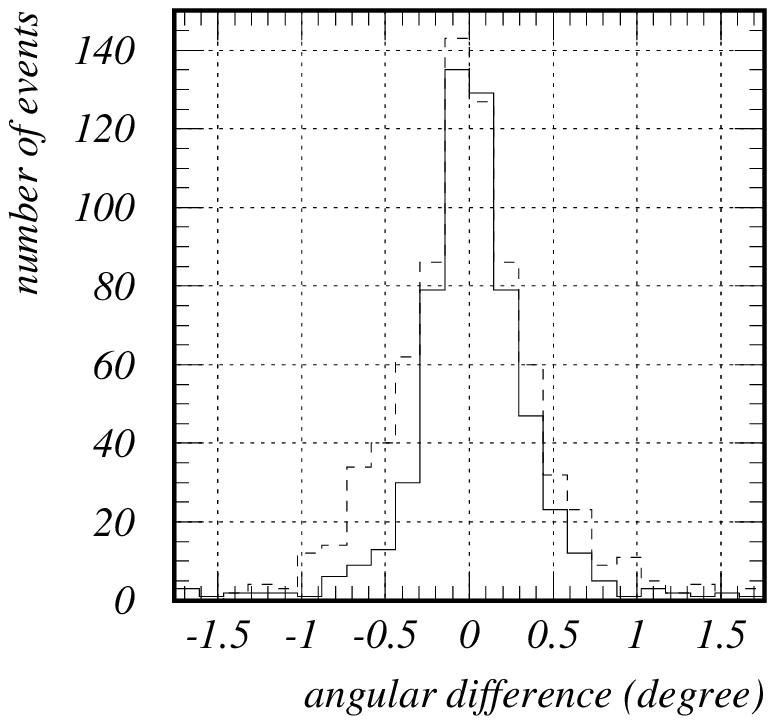}
\caption{\it Angular difference between real and derived directions
for large zenith-angle observations. The solid
histogram was obtained from
the north-south setup and the dashed 
histogram from the east-west setup.}
\label{ns}
\end{figure}

In the case of a largely inclined shower, the geomagnetic effect 
becomes comparable with the intrinsic shower-image size
in the direction perpendicular to the magnetic field.
In that direction, a deterioration of the angular resolution
is expected \cite{ref:mag}.
This means that a difference is introduced
whether the telescopes are separated east-west or north-south.
The difference in the angular resolutions is shown in Figure~\ref{ns}.
Clearly, an east-west setup is better than north-south.
The second CANGAROO-III telescope  will be built
at a position 100\,m West of
the existing CANGAROO-II telescope.

\begin{figure}
\includegraphics[width=11.0cm]{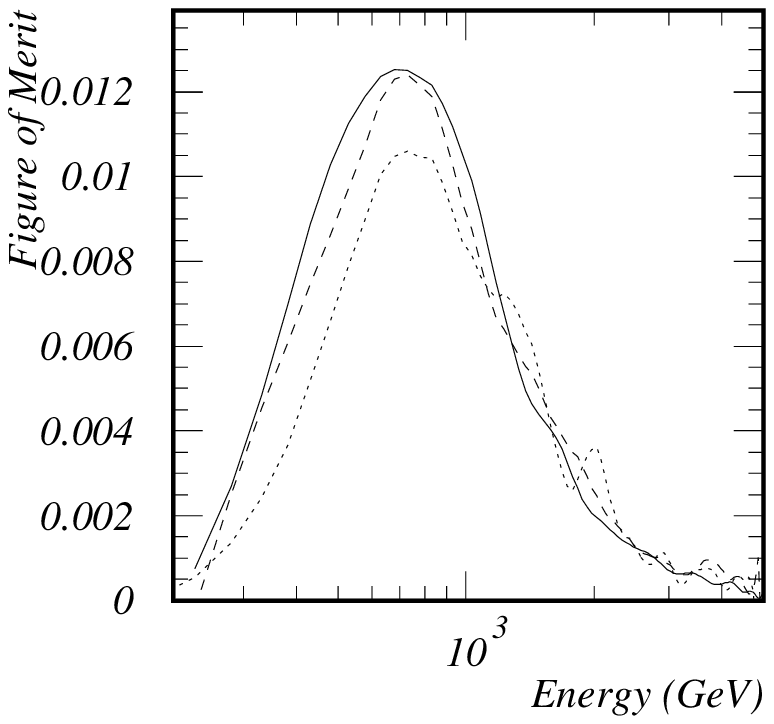}
\caption{\it Figure of merit versus the energies and spacings for 
large zenith angle, stereo mode observations.
The solid line is for 60\,m spacing, the dashed 100\,m, and the dotted 140\,m.
}
\label{fom}
\end{figure}

Finally, we demonstrate the feasibility of the large zenith angle
stereoscopic observations in the case of the east-west setup.
The FOM is shown in Figure~\ref{fom}.
Again, the spacings between 60 and 140\,m are all acceptable. 
The FOM for the 100\,m spacing is located near the maximum.
The energy threshold of this mode at the zenith angle of 55$^\circ$ 
was obtained to be 700\,GeV.
At 700\,GeV, the angular resolution obtained was 0.43$^\circ$
per shower. The effective area was $1.2\times 10^5 m^2$
at 700\,GeV.

\section{Discussion}
Considering the values obtained by this study, we determined the design
of the second CANGAROO-III telescope.
The location of the second telescope is 100\,m west from the present
CANGAROO-II telescope.
The third and fourth telescopes will be located north and south from the
baseline of the first two telescopes, i.e., at the corners of
a diamond shape.
The camera pixel spacing was determined to be 0.168$^\circ$ using
3/4$''$ PMTs. A hexagonal alignment was adopted.
The total field of view is a hexagonal shape approximately 4$^\circ$ across.
The second telescope is scheduled to be built in 2001.

\section{Conclusion}
A design study of the CANGAROO-III telescope system was 
carried out using the Monte Carlo technique. 
The optimized pixel spacing is 0.168$^\circ$ with
use of 3/4~inch PMT's.
This was carried out for gamma-rays from zenith positions.
The telescope spacing for stereoscopic observations was optimized to be
100\,m. The energy threshold at zenith was determined to be 200\,GeV 
with an effective area of $2.4\times 10^4m^2$. 
The angular resolution per shower 
for stereo observations was 0.2$^\circ$ at 200\,GeV, event by event bases.
With this advantage, we can analyze profiles of the broad gamma-ray sources.
For the monocular mode of the two-telescope system,
the effective area was determined to be $1.8\times 10^4m^2$ with a
threshold energy of 100\,GeV. Low elevation observations, at a zenith angle
of 55$^\circ$, were investigated and an effective area of
$1.2\times 10^5m^2$ with an energy threshold of 700\,GeV obtained.
The angular resolution for stereo observations was 0.43$^\circ$ per shower
at 700\,GeV.
CANGAROO-III will be constructed using these design parameters.

\section*{Acknowledgement}

This work was supported by the Center of Excellence, 
and a
Grant-in-Aid for Scientific
Research by the Japan Ministry of Education, Science, Sports and
Culture.

\end{document}